\documentclass[sigconf]{acmart}
\AtBeginDocument{%
  \providecommand\BibTeX{{%
    \normalfont B\kern-0.5em{\scshape i\kern-0.25em b}\kern-0.8em\TeX}}}

\usepackage{hyperref}
\usepackage{float}

\setcopyright{acmcopyright}
\copyrightyear{2022} 
\acmYear{2022} 
\setcopyright{acmlicensed}\acmConference[MM '22]{Proceedings of the 30th ACM International Conference on Multimedia}{October 10--14, 2022}{Lisboa, Portugal}
\acmBooktitle{Proceedings of the 30th ACM International Conference on Multimedia (MM '22), October 10--14, 2022, Lisboa, Portugal}
\acmPrice{15.00}
\acmDOI{10.1145/3503161.3548379}
\acmISBN{978-1-4503-9203-7/22/10}

\begin{document}

\title{Improving Meeting Inclusiveness using Speech Interruption Analysis}



\author{Szu-Wei Fu}
\email{SzuWeiFu@microsoft.com}
\affiliation{%
  \institution{Microsoft}
  \streetaddress{1 Microsoft Way}
  \city{Redmond}
  \state{WA}
  \country{USA}
  \postcode{98045}
}

\author{Yaran Fan}
\email{Yaran.Fan@microsoft.com}
\affiliation{%
  \institution{Microsoft}
  \streetaddress{1 Microsoft Way}
  \city{Redmond}
  \state{WA}
  \country{USA}
  \postcode{98045}
}

\author{Yasaman Hosseinkashi}
\email{yahossei@microsoft.com}
\affiliation{%
  \institution{Microsoft}
  \streetaddress{1 Microsoft Way}
  \city{Redmond}
  \state{WA}
  \country{USA}
  \postcode{98045}
}

\author{Jayant Gupchup}
\email{Jayant.Gupchup@microsoft.com}
\affiliation{%
  \institution{Microsoft}
  \streetaddress{1 Microsoft Way}
  \city{Redmond}
  \state{WA}
  \country{USA}
  \postcode{98045}
}

\author{Ross Cutler}
\email{Ross.Cutler@microsoft.com}
\affiliation{%
  \institution{Microsoft}
  \streetaddress{1 Microsoft Way}
  \city{Redmond}
  \state{WA}
  \country{USA}
  \postcode{98045}
}

\renewcommand{\shortauthors}{Szu-Wei Fu et al.} 

\begin{abstract}

Meetings are a pervasive method of communication within all types of companies and organizations, and using remote collaboration systems to conduct meetings has increased dramatically since the COVID-19 pandemic. However, not all meetings are inclusive, especially in terms of the participation rates among attendees. In a recent large-scale survey conducted at Microsoft, the top suggestion given by meeting participants for improving inclusiveness is to improve the ability of remote participants to interrupt and acquire the floor during meetings. We show that the use of the virtual raise hand (VRH) feature can lead to an increase in predicted meeting inclusiveness at Microsoft. One challenge is that VRH is used in less than $1\%$ of all meetings. In order to drive adoption of its usage to improve inclusiveness (and participation), we present a machine learning-based system that predicts when a meeting participant attempts to obtain the floor, but fails to interrupt (termed a `failed interruption'). This prediction can be used to nudge the user to raise their virtual hand within the meeting. We believe this is the first failed speech interruption detector, and the performance on a realistic test set has an area under curve (AUC) of 0.95 with a true positive rate (TPR) of $50\%$ at a false positive rate (FPR) of $<1\%$. To our knowledge, this is also the first dataset of interruption categories (including the failed interruption category) for remote meetings. Finally, we believe this is the first such system designed to improve meeting inclusiveness through speech interruption analysis and active intervention. 

\end{abstract}



\maketitle

\section{Introduction}
Computer-mediated communication (CMC) systems, especially video conferencing systems, have transformed how companies and organizations conduct meetings. In the past two years since COVID-19 began, the use of such systems has dramatically increased as many people were forced to work from home. A common goal for these systems is to facilitate the most effective meetings possible \cite{cutler2021meeting}. A more recent goal is to also facilitate the most inclusive meetings possible, where everyone feels included in the meeting. A recent study found that the top user suggestion for improving meeting inclusiveness is to better allow the remote meeting participant to interrupt and speak in the meeting \cite{cutler2021meeting}. Using meeting telemetry, that same study shows that meetings are much more likely to be inclusive when users participate and that users are much less likely to participate if they are remote. 

\begin{figure}
 \centering
 \includegraphics[width=\linewidth]{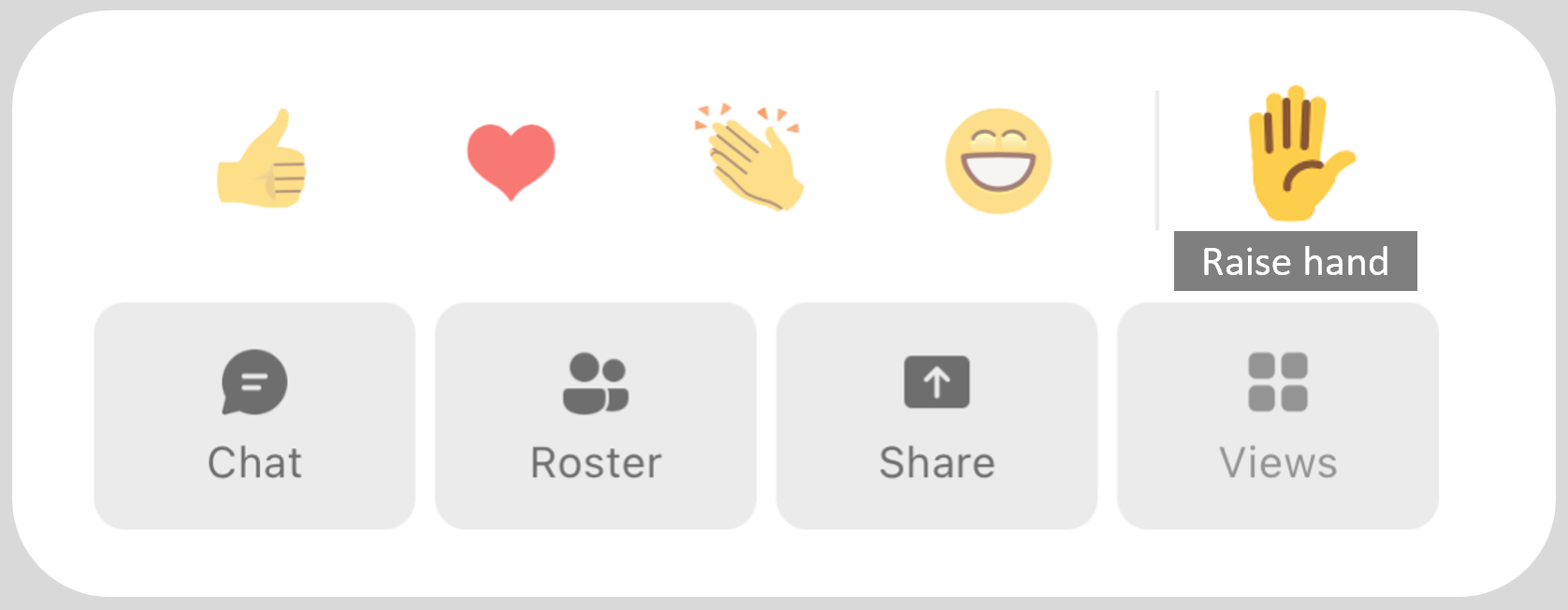}
 \caption{The virtual raise hands (VRH) option present in Microsoft Teams.}  
 \label{fig:vrh}
\end{figure}

To help improve meeting inclusiveness, we address the top issue of improving the ability of remote participants to interrupt and speak in a meeting. While most remote collaboration systems include a VRH feature to help solve this issue (see Figure \ref{fig:vrh}), in Microsoft Teams we found it is used in less than $1\%$ of the meetings. We show that the VRH feature, when used, is effective at improving meeting inclusiveness. We describe a machine learning-based speech analysis model that detects failed interruptions, i.e., when someone tries to speak but doesn't get the floor. This classifier can then be used to nudge users to raise their virtual hand and then speak or notify the meeting facilitator to help ensure the participant gets a chance to speak up in the meeting. 
One of the biggest challenges in building such a classifier is the lack of a labeled and representative dataset. We used 250 hours of meeting data from publicly available sources and created 90 hours of new data to build a representative train and test set of 40,000 clips. These meetings were segmented into audio clips containing speech overlaps. In order to label these clips accurately, we developed a crowdsourced labeling system. We validated the accuracy of the labeling thoroughly, a critical step for ensuring the model achieves the desired accuracy. To the best of our knowledge, this is the first labeled dataset that categorizes the different types of interruptions in meetings.

Production CMC systems with millions of users have very strict accuracy requirements to ensure a smooth user experience. From our production data, we observed 40 interruptions for typical half-hour meetings with 4 participants. We established the FPR of our detector needed to be $\leq$ 1\%; this ensures that an average user would see one false positive every 10 half-hour meetings. Furthermore, we found that failed interruptions represent 10\% of speech overlaps so setting our TPR to $>40\%$ ensures that we capture at least two failed interruptions per meeting on average.  Meeting such a strict criterion required us to evaluate several architectural choices. Failed interruptions are usually short speech utterances. The core challenge is to disambiguate such events with other short speech utterances such as agreements (e.g., ``yeah'', ``makes sense'') and acknowledgments (e.g., ``uh-huh'', ``hmmm''). Given the small number of labeled clips (40K), it would be impossible to train a model from scratch. Therefore we leveraged the recent advances in self-supervised learning (SSL) representations that are learned on tens of thousands of hours of unlabeled data (\cite{baevski_wav2vec_2020}, WavLM \cite{chen_wavlm_2021}).

Our contributions in this paper are:

\begin{itemize}
    \item We frame the problem of detecting failed interruptions and create the first known dataset for categorizing interruptions in remote meetings with a labeling accuracy of 95\%.
    \item We provide the first predictor of failed speech interruption we are aware of. The failed speech interruption predictor achieves an AUC of 0.95 with a TPR > 50\% and FPR < 1\% in a realistic test set, which is good enough for a production CMC system.
    \item We believe this is the first functional system designed to improve meeting inclusiveness through speech interruption analysis.
\end{itemize}

In Section \ref{sec:related_work} we review related work in this area. In Section \ref{sec:hand_raise} we show that when the VRH feature in Microsoft Teams is used, it statistically increases inclusiveness. Section \ref{sec:dataset}
describes the training dataset and test set used by our model, and Section \ref{sec:model} describes the model architecture and training. Section \ref{sec:results} and  \ref{sec:applications}  give the results and some possible applications of the model, respectively. We provide conclusions and future work in Section \ref{sec:conclusions}.

\section{Related work}
\label{sec:related_work}
Characterizing the conversational process is a well-studied topic. Schegloff presents a detailed account of the differences between ``turn-taking'' and ``interruptions'' \cite{schegloff_overlapping_2000}. This work defines an interruption and cites clear examples of what should not be considered interruptions (e.g., signaling the current speaker to continue, gestures such as laughter, annoyance). A formal structure for organized turn-taking is presented in \cite{sacks_simplest_1978}. This work defines the notion of a ``false start'' as turn-taking violations that need to be repaired. These repair rules are intended to address the intent to acquire the floor. However, the work assumes that the repair mechanisms will be followed (which isn't observed in practice). Margariti et al.~present a quantitative analysis of turn-taking experienced in CMC systems \cite{margariti_automated_2022}. They find that if a single speech overlap occurs, the odds of an overtake are roughly the same as the original speaker keeping the turn. However, when the same interrupter attempts multiple times, it is far more likely that an overtake will occur. The work does not differentiate between a failed interruption versus an interjection (also known as a ``backchannel''); both result in a failed overtake.

Sellen studies the impact on turn-taking and interruptions using CMC systems (audio-only and video) compared to same-room conversations \cite{sellen_remote_1995}. Sellen's results show that simultaneous starts (leading to failed interruptions) occurred at the same rate in same-room conditions as they did in technology-mediated solutions. The work also states that a turn (or floor) has been acquired if a speaker is not interrupted for more than 1.5 seconds, a helpful quantitative definition used in our labeling process. Sellen's work, however, assumes good network and device conditions. The effect of network latency on ``conversational interactivity'' in VoIP systems is studied by Hammer et al.~\cite{hammer_well-tempered_2005}. This work introduces the notion of active and passive interruptions. An active interruption is an intentional interruption whereas a passive interruption is unintended and occurs due to the delayed effect of hearing the remote speaker. The authors report that delay impacts passive interruption far more significantly than an active interruption. Every 100ms delay leads to a $15\%$ relative increase in passive interruption rate. Echo suppression artifacts from double talk scenarios lead to attenuation leading to challenges in the ability to interrupt in CMC systems \cite{blind}. 

International Computer Science Institute (ICSI) and AMI multi-party meeting datasets provide a good starting point (ICSI \cite{janin_icsi_2003}, AMI \cite{carletta_ami_2006}) for analysing meeting interruptions. The ICSI dataset contains 75 real meetings and 72 hours of speech. All the meetings have a mixed audio file and one file for each speaker (headset, open microphone, etc.). It comes with detailed word-level transcriptions, and annotations capturing interruptions along with backchannels. AMI is a 100-hour meeting corpus recording in three rooms with different acoustic properties using non-native speakers. Speakers in ICSI and AMI are wearing headsets while being present in the same room. While each speaker is represented as a unique channel, cross-talk between speakers is not eliminated as the microphone associated with one participant is picking up speech from another participant. Remote-only meetings will not have this same issue. To address this bias, we collected our own data using our conferencing solution with remote-only participants.  As a result, we primarily use these data sources to augment our training data; we do not use them to construct our test set as our typical remote-only meeting scenarios do not have cross-talk. These datasets do not contain labels for failed interruptions. To the best of our knowledge, there does not exist a labeled audio dataset that captures the failed interruption category. In this work, we created detailed instructions for crowd-workers to categorize the different interruption types (including failed interruptions).

Detecting speech overlaps in VoIP applications is primarily done through voice activity detectors (VADs) \cite{braun_training_2021, tan_rvad_2020, kim_voice_2018, boakye2008overlapped}. Baron used a prosody-based approach to predict interruptions and ``jump-in'' points. Pitch and pause features significantly outperformed language models in predicting jump-in points \cite{baron_prosody-based_2002}. Makhervaks et al.~build a model for detecting ``hotspots'' (i.e., regions of high interest) during meetings \cite{makhervaks_combining_2020}. The model comprises features derived from speech activity, language model embedding and prosody.  Acknowledgements (or agreements) such as ``hmm'', ``yeah'' are commonly referred to as ``Backchannels'' in the literature \cite{morency_predicting_2008, ruede_enhancing_2017}. In \cite{ruede_enhancing_2017}, Ruede et al.~categorize backchannels using word embeddings. In \cite{kennedy_laughter_2004} Kennedy et al.\ built a support-vector based model for categorizing laughter using  mel-frequency ceptral coefficients (MFCCs). Keyword spotting models represent a special case of accoustic event detection aimed at detecting only a small set of words \cite{sainath_convolutional_2015, sun_max-pooling_2016}. 

To the best of our knowledge, no work has aimed at categorizing the different types of speech overlaps in remote meetings. Disambiguating failed interruptions from backchannels is at the core of this challenge as those two event types have similar acoustic characteristics, and they are very similar in duration. Yang et al.~and Fitzgerald et al.~built models to detect events they term ``disfluencies'' \cite{liu_automatic_2003, fitzgerald_reconstructing_2009}. These include utterance repetitions, revisions, and false starts. In both these works, the assumption is that the disfluencies are repaired, which does not lead to a failed interruption. Creating a model to detect disfluencies (i.e. failed interruptions) is a complex endeavor. This task can benefit from applying self-supervised approaches applied to other similar downstream tasks such as keyword spotting \cite{chen_wavlm_2021}. There is a growing body of SSL representations (embeddings) trained from raw audio data (Wav2Vec \cite{baevski_wav2vec_2020}, HuBERT \cite{hsu_hubert_2021}, WavLM \cite{chen_wavlm_2021}). In terms of the downstream task of automatic speech recognition (ASR), each of these SSL approaches has been able to improve on the previous state-of-the-art, demonstrating the effectiveness of these embeddings for extracting language context. In our work, we use these embeddings generated from raw audio as an input to our interruption classifier. 

There is only one work in measuring meeting inclusiveness we are aware of \cite{blind}. The authors conducted a survey of 16K employees (3.3K valid responses) in a large technology company and developed a multivariate model to extract the relationship between meeting effectiveness, inclusion, and their contributing factors. The study showed that 80\% of the meetings in this company were inclusive, showing a significant room for improvement. The model indicates that participation is the main contributor to the perceived meeting inclusiveness, $47\%$ larger than the next important factor. Additionally, the survey results included that the top feature request from participants to improve meeting inclusiveness was a “better ability for remote participants to interrupt.” 

Gender has been studied as a factor in conversations and participation. Eecke et al.~shows that women are more often interrupted in conversations than men, and that men interrupt women more often than they interrupt men \cite{eecke_influence_2016}. Leaper et al.~\cite{leaper_meta-analytic_2007} conducts a meta-analysis that shows men talk more than women in conversations. James et al.~provides a review of the extensive research on this topic of gender bias in conversation \cite{james_understanding_1993}.

\begin{table*}[t]
\caption{Number of Clips by Source and Label}
\label{table_clips}
\begin{tabular}{ccc|ccccc|c}
\hline
Data Source   & Raw Audio Hours & Labeled By   & Backchannel & Failed Interruption & Interruption & Laughter & Other & Total  \\ \hline
AMI           & 100             & Crowdsourced & 8,760       & 2,270               & 5,630        & 1,650    & 150   & 18,450 \\
ICSI          & 72              & Crowdsourced & 4,650       & 1,060               & 3,570        & 50       & 100   & 9,420  \\
Original Data & 77              & Crowdsourced & 3,840       & 1,140               & 2,570        & 1,430    & 200   & 9,180  \\
Original Data & 15              & Expert       & 1,310       & 320                 & 860          & 340      & 150   & 2,970  \\ \hline
All           & 264             & All          & 18,560      & 4,790               & 12,630       & 3,470    & 600   & 40,020 \\ \hline
\end{tabular}
\end{table*}

\section{Analysis of virtual raise hand and inclusiveness}
\label{sec:hand_raise}
The VRH feature is designed to facilitate the participation of remote participants in a popular video-conferencing application. The goal of this section is to quantify the impact of VRH usage in meeting inclusiveness. Here, VRH usage refers to any interaction with the VRH feature during the meeting by any of the participants (either raising a hand or lowering a hand oneself or on behalf of someone else).  The impact is estimated by comparing the meeting inclusiveness with and without VRH usage for comparable sets of meetings.

In our system, VRH usage information is available for all meetings through call technical telemetry. Meeting inclusiveness is measured using a machine learning (ML) model that predicts an inclusiveness score based on the rest of the call telemetry (except for VRH usage). This approach enables a generalizable estimation based on a large sample of real meetings instead of artificial study groups. Next, we will describe the development of ML predictor for inclusiveness score and its application to estimate the impact of VRH.

\subsection{Predictive Model for Inclusiveness}
The initial survey results from \cite{blind} motivated the development of an in-app end-of-call questionnaire that measures user-perceived meeting effectiveness and inclusiveness for a randomly selected set of meetings. The survey includes two 5-scale questions (3 being neutral): 1. How effective was this meeting at achieving its goals? 2. How included did you feel during the meeting? 

The in-app survey produced more than 40K ratings that when joined with meeting telemetry were used to fit a predictive model for meeting inclusiveness and effectiveness. The model consumes 28 engineered features derived from call quality, reliability, meeting size, audio participation (speaking during the call), video or screen share usage, meeting duration, and meeting time. All these features proved to be important in describing the rating variations via a Graphical model reported in \cite{blind}. Similar to \cite{blind}, the in-app data also showed that participation is one of the dominant features to predict and describe inclusiveness. 

The model used for predictive purposes in this work is a lightGBM \cite{ke_lightgbm_2017} binary classifier that generates the probability of a user providing a 4-star or 5-star rating to the inclusiveness question. Performance on the test set is measured via cross-validation with 50 random test-train splits and shows the AUC to be 0.75+/-0.02 for the test set. 

To predict the user ratings for calls with no user rating, we convert the predicted probability to a binary score by applying a threshold that ensures the FPR is not larger than 5\%. FPR is defined as the rate of incorrect classification when the actual user rating for inclusiveness is not 4- or 5-stars. At 5\% FPR, the model has 32\% TPR and can only flag 32\% of 4- or 5-star calls with the available telemetry. Therefore, the predicted baseline of inclusiveness score by the model is much lower than the actual user ratings (if available). However, the model is still sensitive enough to estimate the lift in predicted scores because of using VRH feature. This is discussed in the next section. 

\subsection{Raise Hand Impact on Predicted Meeting Inclusiveness}
We compared the predicted inclusive score between meetings with and without VRH engagement. Predictions are binary values generated by the model described in the previous section. Since VRH is a feature available to all users and is not subject to any controlled experimentation \cite{kempthorne_design_1952}, it is not possible to conduct a causal analysis. Instead we apply Propensity Score stratification \cite{austin_introduction_2011} as a pseudo-experiment method.

Propensity score (PS) methods are effective techniques for estimating causal relations in the absence of controlled experimentation \cite{rosenbaum_reducing_1984, hullsiek_propensity_2002, cochran_effectiveness_1968}. \cite{cochran_effectiveness_1968} showed that if the propensity scores are set properly, they can eliminate more than 90\% of the bias induced by confounding factors. Confounding factors are any attribute of the data that may have a different distribution in the treatment and control samples, and additionally, their impact on the outcome variable can be mixed with the effect of the outcome variable. In this study, the outcome variable is meeting inclusiveness, the treatment group is the set of calls where VRH is used (called VRH calls), and the control group is the set of calls without VRH engagement. 

In this study, the main confounding factors are meeting size, meeting duration, and the choice of media (e.g., video or screen share). For example, VRH is mostly used in meetings with 4 or more participants that are longer than 30 minutes. Therefore, the distribution of meeting size and meeting duration in VRH calls is significantly different from average calls. To reduce this bias, we generate PS values using a logistic regression model that predicts VRH activity as a function of confounding factors. These PS values range between 0 and 1 and are used to generate five equal-sized bins. Within each bin, the distribution of confounding factors is statistically similar and provides comparable sets for estimating the VRH impact. According to \cite{neuhauser_number_2018-1} increasing the number of bins can improve the accuracy of inference but the margin becomes smaller with more than 10 bins. In our analysis, the results were mostly consistent between 5 and 10 bins with minimal gain in the variance of the predicted delta, i.e., a smaller 95\% confidence interval (CI).

Using PS stratification, the impact of VRH on the predicted inclusive score is a 3.4\% absolute increase with 95\% CI of (2.9\%, 3.9\%). This means that for meetings with more than 2 participants, using VRH can improve the predicted inclusive score by an absolute 3.4\% on average at a 95\% confidence level based on the ML model for inclusiveness. 

\section{Dataset}
\label{sec:dataset}
In this section, we detail our efforts to create an accurate train and test set for developing our interruption detector.
We define the speech overlap categories in Section \ref{subsec:categories} and specify our dataset requirements in Section \ref{subsec:data_requirements}. 
Next, we tabulate the data sources used in Section \ref{subsec:sources}. In Section \ref{subsec:labeling}, we outline our labeling process to provide an account of the nuances in labeling such a dataset. Finally, in Section \ref{subsec:train_test}, the train and test sets are described.

\subsection{Speech Overlap Categories}
\label{subsec:categories}
The speech overlap categorization assumes that the floor is held by a speaker. An interrupter is a second participant that speaks to create overlapping speech; they may or may not intend to obtain the floor.
\noindent
\begin{itemize}
    \item \textbf{Backchannel}: A short period of conversation when a speaker conveys attention, understanding, or agreement in the background. The intention is not to obtain the floor.  For example: “yeah”, “Mm-hmm”, "uh-huh".
    \item \textbf{Failed Interruption}: The interrupter attempts to obtain the floor by speaking at the same time as the current speaker, but they fail to obtain the floor.
    \item \textbf{Interruption (or successful interruption)}: The interrupter overlaps with the first speaker before the sentence of the first speaker is complete. The interrupter successfully obtains the floor and the attention of all the other participants. 
    \item \textbf{Laughter}: The interrupter laughs while the first speaker is talking. The interrupter does not get the floor and does not intend to get the floor.
    \item \textbf{Other}: This represents audio overlap scenarios without clear speech content. This could represent overlaps with no intelligible words (e.g., garbled speech, throat clearing), or background noise (e.g., mouse clicks).
\end{itemize}

\subsection{Dataset Requirements}
\label{subsec:data_requirements}
The CMC system (Microsoft Teams) we are studying can separate the audio of each remote participant. As a result, we only considered data sets where the audio of each participant was captured in a separate channel. This design choice simplifies the solution as a mixed channel dataset would require the speakers to first be separated (i.e., speaker diarization). Our scenario represents multi-party business meetings, hence we were interested in conversational dynamics arising from remote meetings with three (3) or more participants. We restrict ourselves to the English language but captured different accents. We prioritized accents from the following locales: United States (US), Great Britain (GB), India (IN), and Germany (DE). These locales were obtained based on countries with the highest product usage. The dataset should have an equal representation of male and female speakers. The age group of the speakers in the dataset is required to be higher than 18 since these conversations represent business meetings. For our current version of the dataset, we restricted ourselves to speakers in low noise conditions with headsets and good networks. In the future, we plan to capture data with varying network (latency), device (e.g., open speakers), and acoustic (e.g., noise) conditions. 

To ensure that we can validate the quality of predictions from our detector, we set a requirement for the accuracy of the labels in the test set to be $95\%$ as measured by the Fleiss' Kappa metric \cite{nichols_putting_2010}.

\subsection{Data Sources}
\label{subsec:sources}
In our development process, we use both qualified publicly available datasets and our original datasets as input data sources.

\subsubsection{Public Data Source}

We use the AMI Meeting Corpus \cite{carletta_ami_2006} and the ICSI Meeting Corpus \cite{janin_icsi_2003} as our public data sources. Both corpora meet our scenario requirements, however, there are a few limitations too. Both corpora were collected with participants sitting in the same room with headsets on. This results in two problems: first, all meetings are face-to-face, which could be different from online communication; second, with people sitting in the same room, it includes cross-talk when people’s microphones pick up other people’s voices. We use these datasets to augment our training data.

\begin{figure*}
 \centering
 \includegraphics[width=\textwidth]{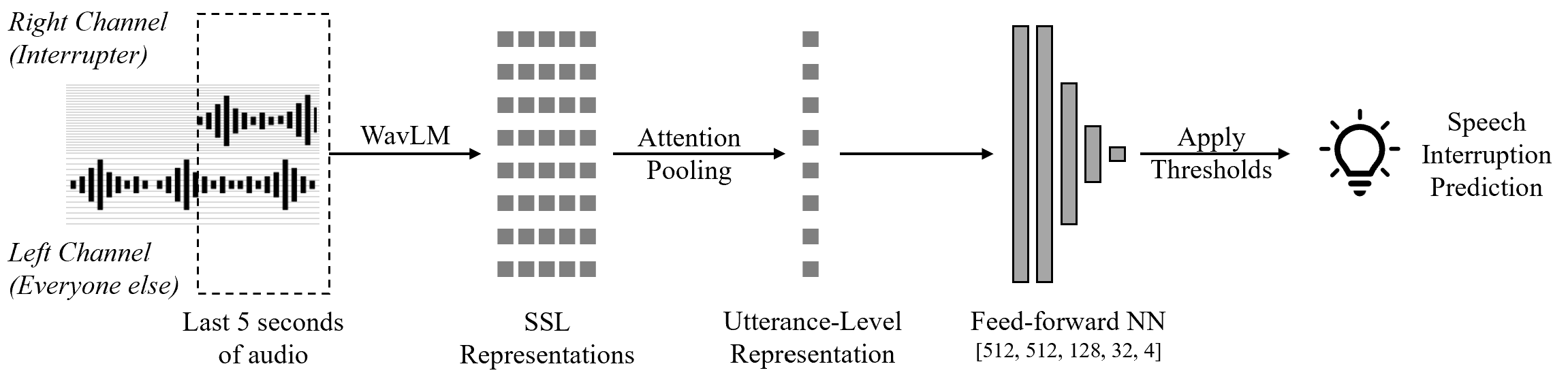}
 \caption{Proposed model structure for speech interruption detection.}  
 \label{fig:model}
\end{figure*}

\subsubsection{Original Data Source} 

To mitigate the challenges from the pubic datasets and create accurate test sets for remote meeting scenarios, we captured data using our conferencing product. Another goal of capturing this data was to improve the dataset diversity. We have collected 90 hours of meeting recordings in 148 meetings. Each meeting has 3 or 4 participants. The participants join remotely using our product, and discuss one or more business topics having some natural contention (e.g., prioritization, product design, etc.). In the data collection stage, we also ensure that we have sufficient speaker diversity for gender, accent, and age. In Table \ref{table_demographic}, We show the detailed demographic information of our original data at meeting-participant level.

\begin{table}[]
\caption{Demographic Information of Original Data Source}
\label{table_demographic}
\begin{tabular}{c|ccc}
\hline
Demographic   Info & Sub-categories & Count & Percentage \\ \hline
Gender             & Female         & 292   & 53\%       \\
                   & Male           & 251   & 46\%       \\
                   & Non-binary     & 7     & 1\%        \\ \hline
Accent             & US             & 256   & 47\%       \\
                   & GB             & 111   & 20\%       \\
                   & IN             & 111   & 20\%       \\
                   & DE             & 63    & 11\%       \\
                   & Other          & 9     & 2\%        \\ \hline
Age Group          & 18-24          & 98    & 18\%       \\
                   & 25-34          & 184   & 34\%       \\
                   & 35-44          & 107   & 19\%       \\
                   & 45+            & 161   & 29\%       \\ \hline
Total              & Total          & 550   & 100\%      \\ \hline
\end{tabular}
\end{table}

\subsection{Labeling Process} 
\label{subsec:labeling}

There are two major steps in creating labeled data: 1. Detect speech overlap using an accurate VAD to create candidate clips, and 2. Label candidate clips using an accurate labeling procedure. These clips were used to create the train and test sets. The training dataset was labeled using a crowdsourcing service \cite{scale}. We had experts label 3,000 clips to create a test set with an accuracy of $\geq 95\%$, and to help validate crowdsourcing label quality using golden sets.  

\subsubsection{Candidate Clips} 
\label{data_clips}

For each meeting, we scan each channel with an accurate VAD \cite{braun_training_2021} to locate the timestamps where speech overlap occurs. Given that we don’t want to trigger our feature too often, especially in active discussions, we also only keep overlaps in which the interrupter has been silent for at least 3 seconds before jumping in and the interrupter’s utterance needs to be at least 0.3 seconds long. 
Then, for labeling and training purposes, we export a 10-second stereo audio clip for each speech overlap detected. The 5th second of each clip is the start point of the speech overlap. To improve labeling accuracy the clips are created in stereo, with the interrupter’s audio on the right channel, while all the other audios are merged into the left channel. 

\subsubsection{Crowdsourcing Labeling}
\label{data_labeling}

We provided detailed instructions to the crowd-workers for their labeling task. The same instructions were used by experts as a validation of the instructions. These instructions specified set-up instructions (e.g., headset usage), training modules, qualification tests, and golden clips to validate the quality of the expert labeling \cite{naderi_open_2020}. These instructions outlined examples to help disambiguate areas of confusion. One noteworthy challenge was to provide labels for clips that had multiple categories (e.g., a backchannel followed by a failed interruption). We follow the practice from the ImageNet paper to create a hierarchy of categories and represent one category per task \cite{deng_imagenet:_2009}. The hierarchy consisted of splitting the overlaps into two levels. Level 1 comprised on {successful interruption versus no-interruption}. For the no-interruption bucket, the following precedence was followed if multiple categories were present: failed interruption > backchannel > laughter > other. This precedence was based on the cost of misclassification (failed interruptions are sparser and hence get higher precedence). We evaluated the quality of the labels by varying the number of annotators and converged on using 7 unique votes from annotators. 

\subsubsection{Expert Labeling}
We randomly selected 3,000 clips from our original data source and had them labeled by internal experts. In particular, 450 of these clips are labeled by five experts. Multiple rounds of discussions were conducted to ensure complete alignment and refinements of the instructions. Overall, 92\% of these clips reach a 5-out-of-5 agreement and a Fleiss’s Kappa of $95\%$. The rest of the 2,550 clips are labeled by up to two experts. The quality of the 2,550 clips was validated by randomly sampling 150 clips and having all experts label them to ensure we met the accuracy bar. The expert labeling effort was critical in creating an accurate test set and evaluating the crowdsourced labeling system. We iterated on this process over multiple rounds to ensure the labeling process had a 100\% coverage of clips with speech overlap. Among those candidate clips, we achieved an accuracy of $95\%$.

\subsection{Train and Test Sets}
\label{subsec:train_test}
The total number of clips labeled by crowd-workers and experts is shown in Table \ref{table_clips}.

\noindent
\textbf{Test Set}: For the test set, we use high-quality expert-labeled clips. We randomly select 250 clips from each category (not considering other) to create a fixed test set of 1,000 clips.

\noindent
\textbf{Train Set}: The train set is comprised of clips labeled by crowd-workers. We found that the 70\% agreement level (i.e., 5 out of 7 provided the same label) provided better model performance compared to the majority vote. This threshold provided a good trade-off between high label quality and training data volume. We did comparisons on 1,100+ clips with the expert labels as ground truth to arrive at this decision. As a result, we only used clips that could reach the 70\% consensus level.

\section{Model}
\label{sec:model}

The overall model structure is shown in Figure~\ref{fig:model}.
In model training, we only use the audio after the speech overlap starts. Therefore, the model input is the last 5 seconds of the two-channel clips described in Section \ref{data_labeling}. The right channel contains only the voice of the interrupter, and the left channel consists of the mixed voice of all other participants. A feature extractor is then applied to the raw waveform to obtain useful features.  Because SSL representations for speech have achieved state-of-the-art performance on several downstream tasks, in this study, we also extract the high-level embeddings through pretrained-SSL models. 

Before feeding the embeddings to the classifier, a pooling operation along the time axis is used to reduce the input dimension. Here we use attention pooling (AP) \cite{tseng_utilizing_2021} to obtain utterance-level representation $\mathbf{U}\in R^{d \times 1}$ from a sequence of frame-level representations $\mathbf{H}$, where $\mathbf{H}\in R^{d \times M}$ is the embedding extracted from the SSL model, $d$ is the dimension of the feature, and $M$ is the number of frames. The attention weight $\mathbf{Q}\in R^{1 \times M}$ of each frame is first calculated using:

\begin{equation}
\mathbf{Q} = {\rm softmax} (\mathbf{WH})\, 
\end{equation}

\noindent where $\mathbf{W}\in R^{1 \times d}$ can be treated as a template to decide which frame is more important. Finally, utterance-level representation $\mathbf{U}$ can be obtained through weighted sum:

\begin{equation}
\mathbf{U} = \mathbf{H}\mathbf{Q}^T\, 
\end{equation}

Utterance-level representation $\mathbf{U}$ is then fed into a 5-layer feed-forward neural network (also called DNN), with the number of node for each layer as [512, 512, 128, 32, 4] and LeakyReLU \cite{maas_rectifier_2013} as the activation function. Cross entropy is applied as a loss function with stochastic gradient descent as an optimizer using a learning rate of 0.0015.

\begin{table*}[ht]
\caption{Classifier results for the \emph{Failed Interruption} class with different input features (Each number is the average result of 10 different runs). Dimension of $H$ and the number of  parameters for the end-to-end model (SSL model + classifier) are also shown (Note that for the dimension $M$, the frame shift for SSL embedding is 20 ms and we use the default setting from torchaudio library for MFCC and spectrogram).}
\label{table_result}
\centering
\begin{tabular}{c|c|c|cc}
\hline
& & & \multicolumn{2}{c}{Failed Interruption:} \\
 \hline
Input Feature  & dimension of $H$ $({d \times M})$ & \# parameters & AUC   &  TPR@ 1\% FPR \\ 
\hline
MFCC     & (2$\times$40) $\times$ 401  & 0.4M & 0.667 & 3.48\%  \\
Spectrogram   & (2$\times$257) $\times$ 313  & 0.4M & 0.736 & 6.24\% \\
openSMILE & (2$\times$88) $\times$ 1 & 0.4M & 0.816 & 11.56\% \\
$Wav2vec2_{Base}$ & (2$\times$768) $\times$ 249 & 95M & 0.925 & 32.28\%                            \\
$HuBERT_{Base}$  & (2$\times$768) $\times$ 249 & 95M & 0.934  & 31.52\%                            \\
$WavLM_{Base+}$  & (2$\times$768) $\times$ 249  & 95M & 0.943 & 37.80\%                            \\
$WavLM_{Large}$  & (2$\times$1024) $\times$ 249 & 316M & 0.949 & 50.93\%                            \\ \hline
\end{tabular}
\end{table*}

\section{Results}
\label{sec:results}
\subsection{Performance Comparison Between Different Input Features and Classifier Model}
\label{sec:performance_comparison}

In this section, we first show the results of different feature extraction methods. In addition to the SSL embeddings, the results of using MFCC,  magnitude spectrogram, and openSMILE acoustic feature \cite{eyben_opensmile_2010} are also presented as baselines (for openSMILE  feature, we didn't apply AP, as it is already an utterance-level representation). Because in this study we care more about the performance of detecting the \emph{failed interruption} class, in Table~\ref{table_result}, both AUC and TPR at 1$\%$ FPR are based on the \emph{failed interruption} class. From the table, it can be observed that the performance of conventional features such as MFCC and magnitude spectrogram are far away from our goal. The reason may be because a suitable feature for this task should be \textbf{speaker/noise independent} \cite{hung_boosting_2022} and contain \textbf{semantic information}. It may be hard to extract useful information and discard useless one from traditional features with limited training data. Comparing the three SSL-based embeddings with base model size (i.e., $Wav2vec2_{Base}$, $HuBERT_{Base}$ and $WavLM_{Base+}$), WavLM performs the best which is consistent with the results shown in the SUPERB benchmark \cite{yang_superb_2021}  for other speech tasks. For these three embeddings, we calculate the weighted sum of the representations from different
transformer layers with learnable weights as the input to the attention-pooling module. We found that this can significantly improve the performance compared to that only using the embedding from the last transformer layer. To further improve the performance, we apply $WavLM_{Large}$ for feature extraction and get a TPR > 50\% at FPR=1\%.

\begin{figure}
 \centering
 \includegraphics[width=1\linewidth]{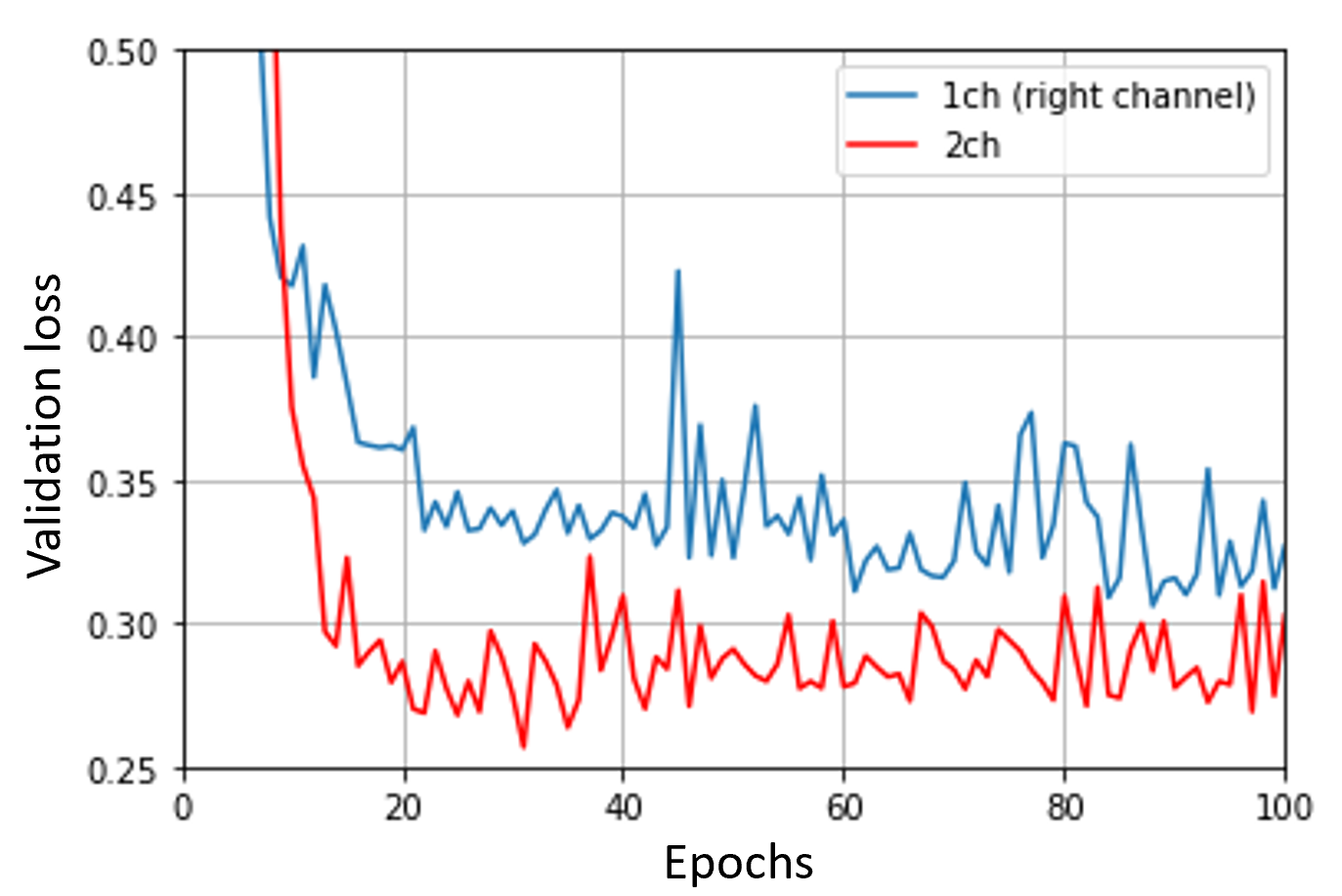}
 \caption{Learning curve with different numbers of input channels}  
 \label{fig:val_loss}
\end{figure}

Next, we fix $WavLM_{Large}$ as our feature extractor and see the performance with different numbers of input channels and classifier models. From Figure ~\ref{fig:val_loss} and the first two rows in Table ~\ref{table_wavlm_large}, we can observe that without the information from the left channel (mixed voice from all other participants), the performance drops seriously, which may be caused by the confusion between failed interruption and successful interruption (will verify this in the next section). We then want to see the effect of AP on the model performance. Removing AP from the downstream task means that the flattened frame-level representations $\mathbf{H}$ are directly fed into the DNN classifier. This not only increases the model size of the classifier significantly but from the third row in Table ~\ref{table_wavlm_large}, the TPR also decreases by 20\%. The reason may be because the input dimension is too large compared to the number of training data.

\begin{table}
\caption{Comparison between different number of input channels and classifier models with $WavLM_{Large}$ as the feature extractor}
\label{table_wavlm_large}
\centering
\begin{tabular}{c|c|cc}
\hline
& & \multicolumn{2}{c}{Failed Interruption:} \\
 \hline
Input channels &classifier model     & AUC   &  TPR@ 1\% FPR \\ 
\hline
2 channels & AP+DNN & 0.949 & 50.93\% \\
right channel & AP+DNN & 0.918
& 37.93\%    \\
2 channels   & DNN    & 0.912 & 31.05\%     \\
                            
\hline
\end{tabular}
\end{table}

\subsection{Confusion Matrix Analysis}
To further analyze the prediction results of $WavLM_{Large}$, we present the confusion matrix in Table ~\ref{table_Confusion_matrix} and Table ~\ref{table_Confusion_matrix_1ch} for two channels and one channel input, respectively. We included another column called `Below 1\% FPR-threshold' to represent the cases where the  
failed interruption should be the predicted class by the $argmax(.)$ function, but the confidence score is smaller than the threshold for a 1\% FPR. From Table ~\ref{table_Confusion_matrix}, it can be observed that the most confusing case for the model is to distinguish between backchannel and failed interruption. We argue that this is because both of them have very similar voice activity patterns (i.e., for the interrupter: a short period of speaking, and for other participants: keep talking). In other words, the model cannot simply tell them by audio energy distribution, it has to infer the intention through semantic information from SSL embeddings. By comparing Table ~\ref{table_Confusion_matrix} and Table ~\ref{table_Confusion_matrix_1ch}, when the ground truth is failed interruption, the model with one channel input is more easily misclassified as a successful interruption, which verifies our assumption made in the previous section that left channel (mixed voice from all other participants) can help the model to distinguish between successful and failed interruption.

\begin{table*}[t]
\caption{Confusion matrix of the prediction results from $WavLM_{Large}$ for \textbf{two} channels input.}
\label{table_Confusion_matrix}
\begin{tabular}{c|cccccc}
\hline
  & & & Predicted \\ \hline
Ground Truth  & Backchannel & \begin{tabular}[c]{@{}c@{}}Failed\\ Interruption\end{tabular} & Interruption & Laughter & \begin{tabular}[c]{@{}c@{}}Below\\ 1\% FPR-Threshold\end{tabular}\\ \hline
Backchannel  &219       &6             &8        &11                &6\\
Failed Interruption  &47     &131            &19         &7               &46\\
Interruption &9       &1           &225         &7                &8\\
Laughter     &37       &0             &1       &212                &0\\ \hline
\end{tabular}
\end{table*}

\begin{table*}[t]
\caption{Confusion matrix of the prediction results from $WavLM_{Large}$ for \textbf{one} channel input.}
\label{table_Confusion_matrix_1ch}
\begin{tabular}{c|cccccc}
\hline
  & & & Predicted \\ \hline
Ground Truth  & Backchannel & \begin{tabular}[c]{@{}c@{}}Failed\\ Interruption\end{tabular} & Interruption & Laughter & \begin{tabular}[c]{@{}c@{}}Below\\ 1\% FPR-Threshold\end{tabular}\\ \hline
Backchannel      &201       &4             &3        &20               &22\\
Failed Interruption  &42      &85            &35        &12               &76\\
Interruption  &8       &4           &207        &11               &20\\
Laughter     &29       &0             &0       &221                &0\\ \hline
\end{tabular}
\end{table*}

\subsection{Relation to Other Speech Tasks}
\begin{figure}
 \centering
 \includegraphics[width=0.9\linewidth]{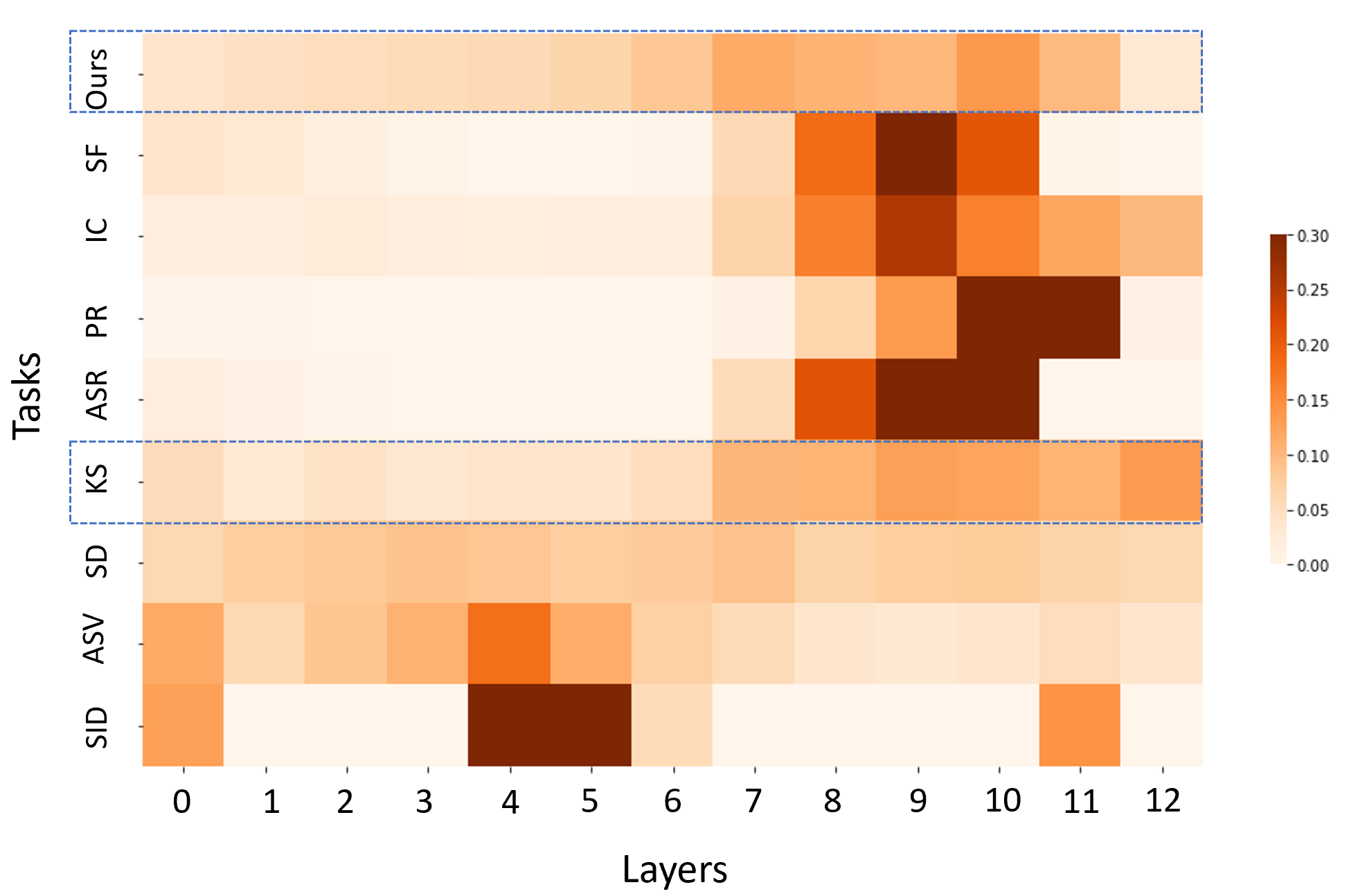}
 \caption{The learned weights from different layers of the $WavLM_{Base+}$ transformer model for different tasks.}
 \label{fig:Weights_layers}
\end{figure}

Speech interruption classification is a relatively new topic. As a result, we want to know its relation to traditional speech tasks. As mentioned in Section ~\ref{sec:performance_comparison}, the SSL-based embeddings come from the weighted sum of different transformer layers with learnable weights. The learned weights can hence give us some information about which layers are more important for a certain task. In Figure~\ref{fig:Weights_layers}, we take the learned layer weights from $WavLM_{Base+}$ as an example to compare our weights with those learned in other speech tasks of the SUPERB benchmark. From the figure, we can observe that the pattern of our learned weights is most similar to the one learned in the Keyword Spotting (KS) task. This implies that the input features used for the two tasks are similar to each other, and they are the most related tasks. We conjecture that this is because when the model tries to distinguish between backchannel and failed interruption, it relies on some mechanism similar to KS (e.g., in the case of backchannel, the interrupter usually says something like: “yeah”, and “Mm-hmm”, etc.)

\section{Applications}
\label{sec:applications}
\begin{figure}
 \centering
 \includegraphics[width=0.9\linewidth]{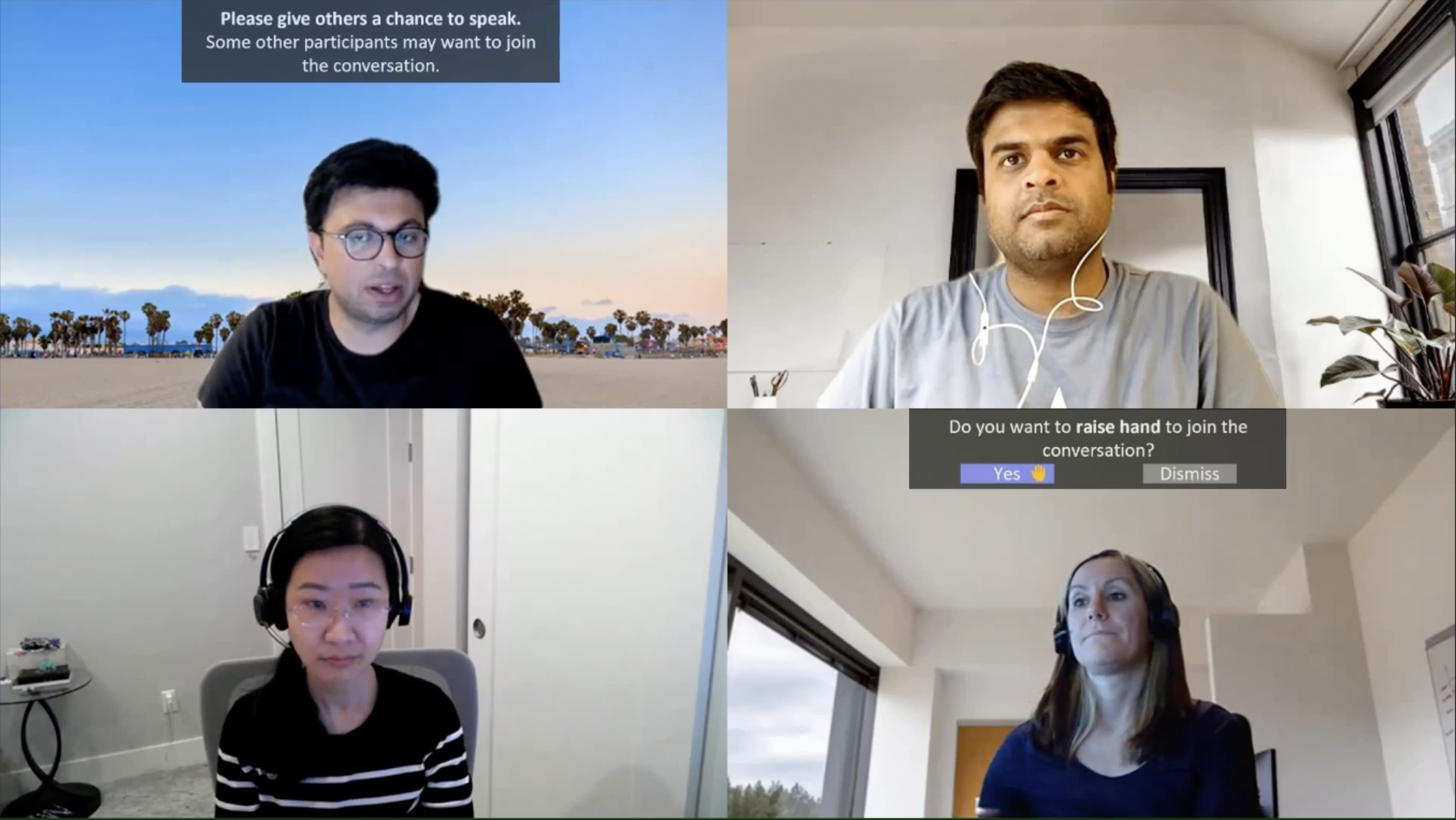}
 \caption{Example application of the failed speech interruption detector}  
 \label{fig:application}
\end{figure}

As an example application of the failed speech interruption detection, we can nudge the failed interrupter to use the VRH feature, as shown in Figure \ref{fig:application}. In addition, we can also remind the speaker who didn't yield that meetings are more inclusive when everyone has a chance to speak, especially if they are repeatably not yielding in that meeting. Finally, the failed speech interruption detections can be logged in the CMC system's calling telemetry and analyzed to further improve understanding of meeting effectiveness and inclusiveness, to conduct AB tests to evaluate new features in the CMC system to reduce failed speech interruptions, and to improve the predictive models of effectiveness and inclusiveness. 

\section{Conclusions}
\label{sec:conclusions}
In this paper, we describe a method to improve meeting inclusiveness through speech overlap analysis. We introduce the challenge of ``failed interruptions'' in remote meetings based on our findings. We created the first accurate labeled dataset to address this challenge in CMC systems. By leveraging recent advances in self-supervised learning representations in speech, we built a detector that achieves a TPR of more than 50\% with a FPR of less than 1\%. The dataset needs to be expanded to support multiple languages and scenarios such as varying network and device conditions. We plan to use this dataset to host a challenge and release the baseline model for stimulating active research on this topic. We are also integrating the model into Microsoft Teams and conducting AB tests to measure the improvement of inclusiveness and overall meeting effectiveness. Finally, we are developing a full-duplex machine learning-based acoustic echo canceller (e.g., see \cite{sridhar2021icassp}) that also helps remote participants interrupt in a meeting and better participate in meetings.

\section*{Acknowledgements}
We acknowledge Juhee Cho, Alex Chzhen, and Chinmaya Madan from the Teams team for their contributions in the model and data infrastructure. The Azure cognition team provided strong guidance with the model development effort. Specifically, we thank Yu Wu, Zhou Chen, and Dimitrios Dimitriadis for the productive discussions.

\clearpage
\bibliographystyle{ACM-Reference-Format}
\bibliography{IC3-AI}

\end{document}